# Concept for a Time-of-Flight Small Angle Neutron Scattering Instrument at the European Spallation Source

S. Jaksch[1,2], D. Martin-Rodriguez[1,2,3], A. Ostermann[4], J. Jestin[5], S. Duarte Pinto[6], W.G. Bouwman[6], J. Uher[7], R. Engels[8], G. Kemmerling[8], R. Hanslik[9], H. Frielinghaus[1,2]

[1]ESS Design Update Programme – Germany, Forschungszentrum Jülich GmbH, Jülich, Germany
[2]Jülich Centre for Neutron Science at Heinz Maier-Leibnitz Zentrum, Forschungszentrum Jülich GmbH, Garching, Germany
[3]Neutron Optics and Shielding Group, European Spallation Source AB, Lund, Sweden
[4]Heinz Maier-Leibnitz Zentrum (MLZ), Technische Universität München, Lichtenbergstr. 1, 85748 Garching, Germany
[5]Laboratoire Léon Brillouin, LLB, CEA—Saclay, Gif sur Yvette cedex, France
[6]Faculty of Applied Sciences, Delft University of Technology, Delft, Netherlands
[7]Amsterdam Scientific Instruments, Amsterdam, Netherlands
[8]Zentralinstitut für Engineering, Elektronik und Analytik (ZEA-2), Forschungszentrum Jülich GmbH, Jülich, Germany
[9]Zentralinstitut für Engineering, Elektronik und Analytik (ZEA-1), Forschungszentrum Jülich GmbH, Jülich, Germany

**Abstract**
A new Small Angle Neutron Scattering instrument is proposed for the European Spallation Source. The pulsed source requires a time-of-flight analysis of the gathered neutrons at the detector. The optimal instrument length is found to be rather large, which allows for a polarizer and a versatile collimation. The polarizer allows for studying magnetic samples and incoherent background subtraction. The wide collimation will host VSANS and SESANS options that increase the resolution of the instrument towards μm and tens of μm, respectively. Two 1m² area detectors will cover a large solid angle simultaneously. The expected gains for this new instrument will lie in the range between 20 and 36, depending on the assessment criteria, when compared to up-to-date reactor based instruments. This will open new perspectives for fast kinetics, weakly scattering samples, and multi-dimensional contrast variation studies.

1. Introduction

The general aim of this SANS concept was to design a versatile instrument for the European Spallation Source (ESS), which allows for a very high dynamic Q range of three orders of magnitude from $10^{-3}$ to ~1 Å$^{-1}$ with a very high Q resolution. At the same time we tried to keep the classical sample sizes of 1x1cm² commonly used in nowadays instruments. The classical sample size allows for highest scattering intensities at still well accepted conditions for many kinetic experiments, for example in stopped-flow cells. In this way, smallest time slices will be possible in kinetic experiments (in most cases well down to the mixing limit of up-to-date stopped flow setups). But also for all other experiments, fast measurements are highly favorable to test a maximum of conditions at a given experimental time. Especially the scientific output of many experiments lies in the variation of parameters that reveal the central mechanism to be studied.

The ESS provides neutron pulses of 2.8ms length at a repetition rate of 14Hz [1], and therefore the ESS is a unique long pulse source, which serves instruments for cold neutrons better than other spallation sources. The optimization of a SANS instrument is still a question that we try to give an answer for within this manuscript.

We argue for a rather long instrument with a collimation of 20m and a detector tube of 20m (Fig. 1). In such long instruments the implementation of neutron polarization does not affect the overall design, as the necessary additional length is negligible. This serves the versatility of the instrument. Polarization allows for studying materials with magnetic domains or fluxlines [2-4], as well as to separate coherent and incoherent background [5]. This separation is highly desirable especially for weakly scattering samples with important details at larger Q, so especially for biological molecules and complexes [6].

In recent years it has become more and more common to investigate samples in situ under a variety of physical conditions [7-8]. To enable the users either to bring custom made sample environments or to install large respectively heavy equipment at sample position we propose a large sample area of up to 3x3 $m^2$. This allows for large magnets [9] or other custom sample environments (cryostats, load frames [10], high temperature stages), and polarization analysis (preferentially with $^3$He) [5].

The central issue for the proposed SANS concept is a wide Q-range with the maximum intensity possible for the smallest Q, because for good statistics the experimental counting times mainly depend on the weakest intensities. The condition of a wide Q-range usually favors rather short instruments [11-12], if the instrument is operated at a spallation source. In the case of short instruments a wide bandwidth of wavelengths can be used, thus at fixed detector geometries, the shorter instruments serve for smaller Q. In contrast, the cold spectrum of the cold source provides the highest intensities at short wavelengths, which usually favors large Q. The underlying concept of this instrument is to employ relatively small wavelengths with high intensities for the small Q by means of a high detector resolution as well as constructing the whole instrument with a relatively large length of 9.5+20+20m (for bender and polarizer/ collimation/detector tube). This concept renders homogenous intensities and therefore comparable statistics for the small and the high Q area. Thus, counting times for both Q areas are in the same time regime.

Another favorable feature of the long instrument is the good wavelength resolution [13]. Here, the spread of the wavelength band is naturally high, and the different wavelengths are much better separated, thus discrimination between different times of flight is more accurate. The different time slices at the detector position allow for wavelength uncertainties of typically 4 to 8%, and can be even reduced to ca. 2% when choosing only large wavelengths. This wavelength resolution can be relaxed if needed in favor of higher intensities by moving the detector set to shorter distances and thus widening the usable wavelength band and therefore higher intensities.

2. Instrument Overview

The proposed SANS instrument has a maximum sample-to-detector distance of 20m and a maximum collimation length of 20m. According to our argumentation this serves for highest intensities at the low Q end and for balanced statistics over the whole Q-range.

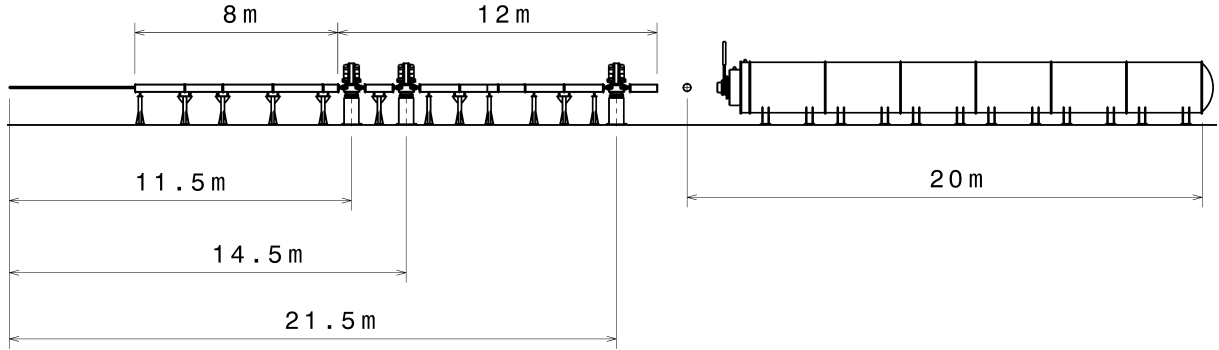

**Fig. 1**: Layout of SKADI after the beam preparation. From left to right: Slim and wide collimation with choppers at 11.5 , 14.5 and 21.5m, sample position and the detector tube. The slim collimation with a length of 8m is located before the first chopper, while the wide collimation with a length of 12m follows after that. Total sample detector distance is 20m.

Including the beam preparation of 9.5m the overall instrument length is 50m (+2m for moving detector tank), measured from the source to the end of the detector tank. Fig. 1 shows the layout of the instrument.

The cold source is designed by the ESS (in terms of simulations and in reality) and operates at 19K [14]. At the shortest distance of 2m, the bender is placed with a length of 4m. Then, a polarizer with radio frequency spin flipper will follow. After this, the collimation of 20m length is placed. The first 8m will be narrow for reasons of angular space between different instruments. After that, the last 12m will have a wider housing for more equipment. Currently, focusing slits for VSANS and a SESANS option are under consideration directed towards larger length scales (down to ca. $5 \times 10^{-4}$ and several $10^{-6}$Å$^{-1}$). The evacuated detector tube will host two 1m$^2$ area detectors, the first one of which has a 20x20cm$^2$ aperture. At the moment we are investigating different technologies for the detector. We aim to achieve high spatial resolution of around 3x3 mm$^2$ with $^6$Li-scintillation converters and subsequent photomultipliers.

As for all time-of-flight instruments, the concepts base on the general equation:

$$\Delta t = \frac{m_n}{h} \Delta \lambda\, L \qquad (1)$$

with $\Delta t$ being the time difference considered, $\Delta\lambda$ the total wavelength or wavelength bandwidth, and L the distance between the events of the generation in the cold source and the detection on the detector. The constant $h/m_n$ (Planck constant divided by neutron mass) takes the value $3.956 \times 10^{-7}$ Js/kg.

3. Estimation of the ideal instrument length

A time-of-flight instrument ideally makes use of the pulse with a rather broad bandwidth in order to maximize the integral intensity. On the other hand, there are limitations for length, and more important, the statistics of the collected intensities of the different wavelengths must stay comparable. So the question is how to find the ideal length for a SANS instrument at a pulsed source. For the ESS, the pulses have a length of 2.8ms at a repetition rate of 14Hz. An analytic estimation is developed in the following:

The spectral intensity from the cold source decays with $\lambda^{-5}$ in the limit of long wavelengths and describes the reality quite well for wavelengths above 3Å for cold source at 19K [14]. The spectrum can be derived from the Boltzmann distribution when changing the dependence from velocities to wavelength and taking absorption by windows into account. The collection of intensities at constant Q on an area detector arises from concentric circles, the circumference of which grows with $\lambda$. Another factor is the transmission T of the sample that usually is described with $T = \exp(-c\lambda^2) \approx 1-c\lambda^2$. The constant c arises from the total coherent scattering cross section, that growths with $\lambda^2$, and we assume a realistic value of $c=0.002$Å$^{-2}$ for typical 1mm thick samples. The incoherent scattering is rather wavelength independent for cold neutrons. The integral intensity for the pulse then is proportional to:

$$I \propto \int_{\lambda_0}^{\lambda_1} d\lambda \cdot \lambda^{-5} \cdot \lambda \cdot (1 - c\lambda^2) = \left[ -\frac{1}{3}\lambda^{-3} + \frac{c}{\lambda} \right]_{\lambda_0}^{\lambda_1} \qquad (2)$$

The limiting wavelengths $\lambda_0$ and $\lambda_1$ contain the full bandwidth also in the center of the detector. However, in the center of the detector at a fixed low angle, low Q are connected to larger $\lambda_0$. Contrarily, at the outer detector rim at a fixed angle high Q are connected to smaller $\lambda_1$. So, the question is, how comparable statistics are achieved at the detector rims without neglecting too many useful neutrons. In Figure 2 we compare the normalized roots of the intensity (for reasons of statistics) as a function of $\lambda_0$ and $\lambda_1$ with the other being at the full bandwidth limit. We compare an instrument length of 17 and 50m with corresponding bandwidths of 16.5 and 5.5Å. The minimal wavelength was chosen to be 3Å for reasons of Double-Bragg-Scattering of window materials.

For the outer rims we see, that the statistics stay sufficiently high for relatively small $\lambda_1$. This means, that the contribution to the counting statistics is higher for the short wavelengths, and the fraction of the pulse is not strongly influenced from the high wavelength side. In both cases the lower limit of the usable wavelength band is $\lambda_1 \approx 4\text{Å}$, which means large Q are barely cut by the fading wavelength band at the outer rim.

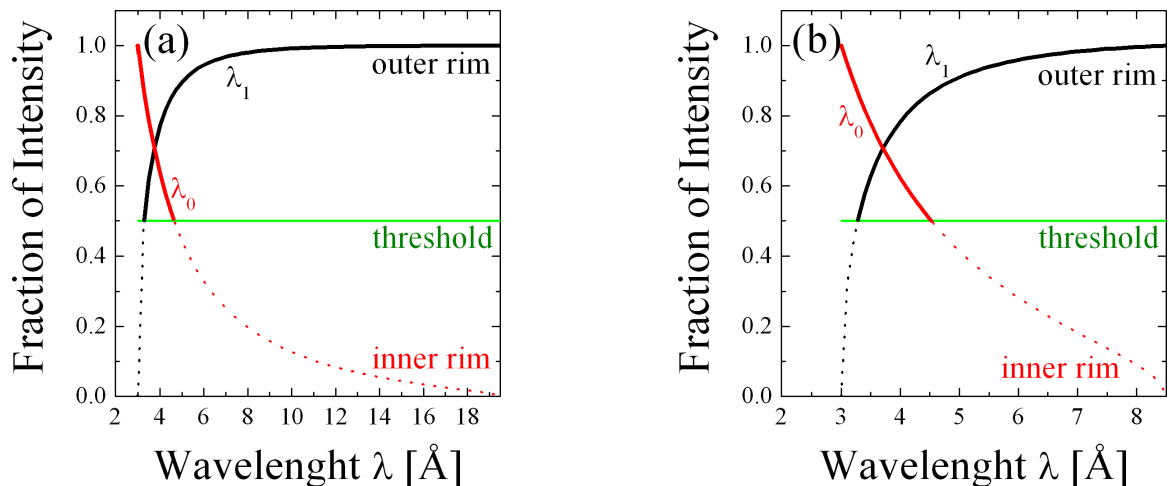

**Fig. 2:** The normalized square root of the intensity for fading wavelength bands at the detector rims as a function of $\lambda_0$ (red, in the case of the inner detector rim) and $\lambda_1$ (black, in the case of the outer detector rim) while the other limit is always the full bandwidth limit. For a 17m long instrument (a) at the outer rim (black) and the inner rim (red) with the threshold (green). For a 55m long instrument (b) the wavelength band is better balanced.

For the inner rims we see similarly, that the short wavelengths contribute stronger, but this results in a fast decay of the normalized root intensity with $\lambda_0$. The acceptable wavelength band does not strongly depend on the instrument length, i.e. $\lambda_0$ of ca. 5Å is acceptable in either case. But for the longer instrument the fraction of the overall band that is affected by a cutoff at a higher $\lambda_0$ is smaller. So, if we aim at small Q, intensities are better balanced in case of a longer instrument, where the cutoff for $\lambda_0$ is at roughly 30% of the bandwidth instead of about 9%. This means, that the typical high intensities of the short wavelength neutrons serve for good statistics in a short time, and experimental mixing of considerably longer wavelengths with much lower intensities does not give an advantage here, because anyway the statistics for low Q at the inner rim of the detector is already dominated by the smaller wavelengths.

The whole consideration is quite robust against variations of the model, such as the exponent -5 for the neutron spectrum. Qualitatively, longer instruments always make better balanced use of the shorter bandwidth.

The impact of the instrument length on the achieved Q-range can be illustrated by computer simulations [15-16]. The shorter instrument allows for a wider wavelength band and vice versa. So the short instrument serves the lowest Q by rather long wavelengths with lower intensities, while the long instrument achieves this result at higher intensities. A summary of these simulations is found in Fig. 3a. The beam preparation (bender, polarizer) takes place in all cases at the first 9m. Equal lengths for the collimation and detector tube were assumed. The minimum wavelength was 4Å, and wavelength bands of 10, 5.5, and 4Å were obtained for the 10+10, 20+20, and 30+30m instruments. All instrument show a $Q^2$ behavior in the central Q-range due to the growing solid angle at constant Q and the Q-binning on logarithmic scale. The instruments have the same entrance aperture, and so the divergence and central intensity decays with a factor of 1, 0.25, 0.11 with growing instrument length. For the 10+10m instrument, the intensity decays constantly towards lower Q from $Q<2\times10^{-3}\text{Å}^{-1}$, and at $Q=10^{-3}\text{Å}^{-1}$ the intensity is not adapted to the overall intensity of the experiment anymore. The 20+20m instrument deviates only slightly from the general $Q^2$-trend at $Q=10^{-3}\text{Å}^{-1}$, and 30+30m instrument even reaches out for smaller Q. The conclusion of these simulations is that the 20+20m instrument performs best for the minimum $Q=10^{-3}\text{Å}^{-1}$ that we believe is necessary for standard SANS instruments. The relaxed resolutions of the longer instruments can also be achieved to obtain highly comparable intensities in the central Q-range.

While the central intensities differ strongly for the strongest collimations (smallest reasonable divergences), the typical intensities of the different instruments are compared best at same collimation settings (as done in Fig. 3b). So, the collimation distance was set to 5m for each instrument, while the longer collimations of each instrument were bridged by neutron guides for the remaining length individually. Here, we see that the short instrument serves the smaller Q better with the high wavelengths, but the low-Q drop stays always very steep, and the corresponding data has to be discarded.

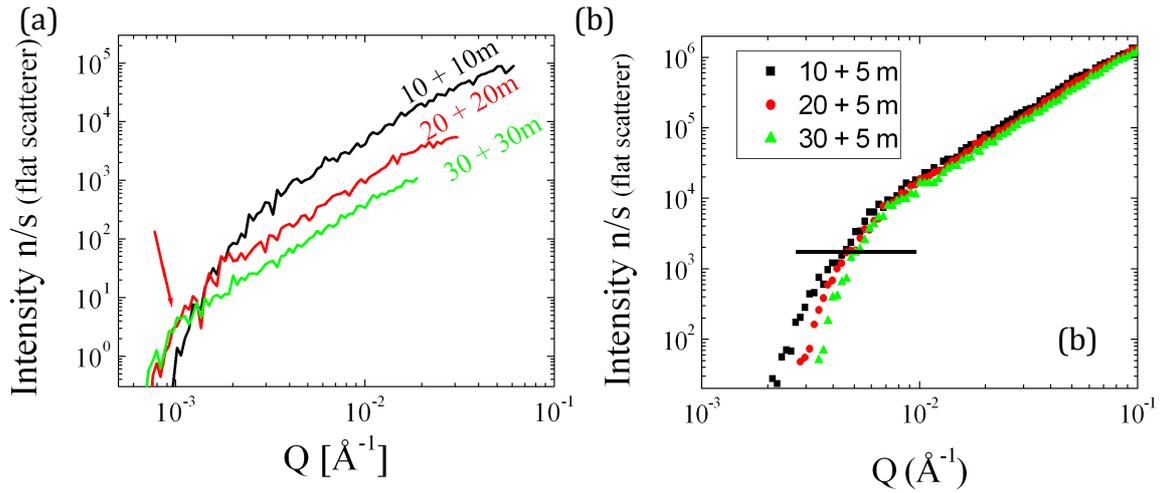

**Fig. 3:** (a) Simulated intensities of SANS instruments with different collimation and detector lengths. The sample was assumed to scatter incoherently flat. The initial beam preparation was assumed to have the same length of 9m in all cases. The increase of the intensity with $Q^2$ arises from the circumference of the detection area (radial addition) and the binning of the intensity ($\Delta Q \sim Q$) on a logarithmic scale. (b) The intensity comparison at the same divergence (5m collimation) and same detector distance (5m). The larger wavelengths of the short instrument slightly elevate the intensity, and slow the low-Q drop a little down. The low Q cutoff would be nearly the same for all instruments (horizontal line).

4. Beam delivery

The beam delivery contains a bender and a polarizer. The minimum distance of guide elements from the source was 2m. Here, a multichannel bender with a length of 4m was placed, followed by the optional transmission polarizer with 2-cavities. All total cross sections were 3x3cm$^2$ (also of the following collimation guides). The default m-coating for straight guides was 1.

4.1 Bender

In a spallation source a direct line of sight between source and sample position is to be avoided, as otherwise high energy neutrons might reach the sample. Similarly a reflection on a neutron guide might serve as a secondary source, so also a direct line of sight between the first reflection and the sample is to be avoided. A bent neutron guide, a bender, achieves this. The bender was a 4-channel bender with a radius of 78m and 4m length. The inner and outer surfaces were coated with m=4. The overall cross section was 3x3cm$^2$. This setup serves well for neutrons of wavelengths above 2Å. The concept was transferred from ESS guidelines, and is not changed throughout the manuscript.

There are clear hints that single cavity benders can transport the desired high brilliance of the source better to the sample (at low divergencies), but the line-of-sight conditions for these constructions are relaxed a little. So, the definite bender construction must be done in collaboration with the ESS after a clear policy is negotiated.

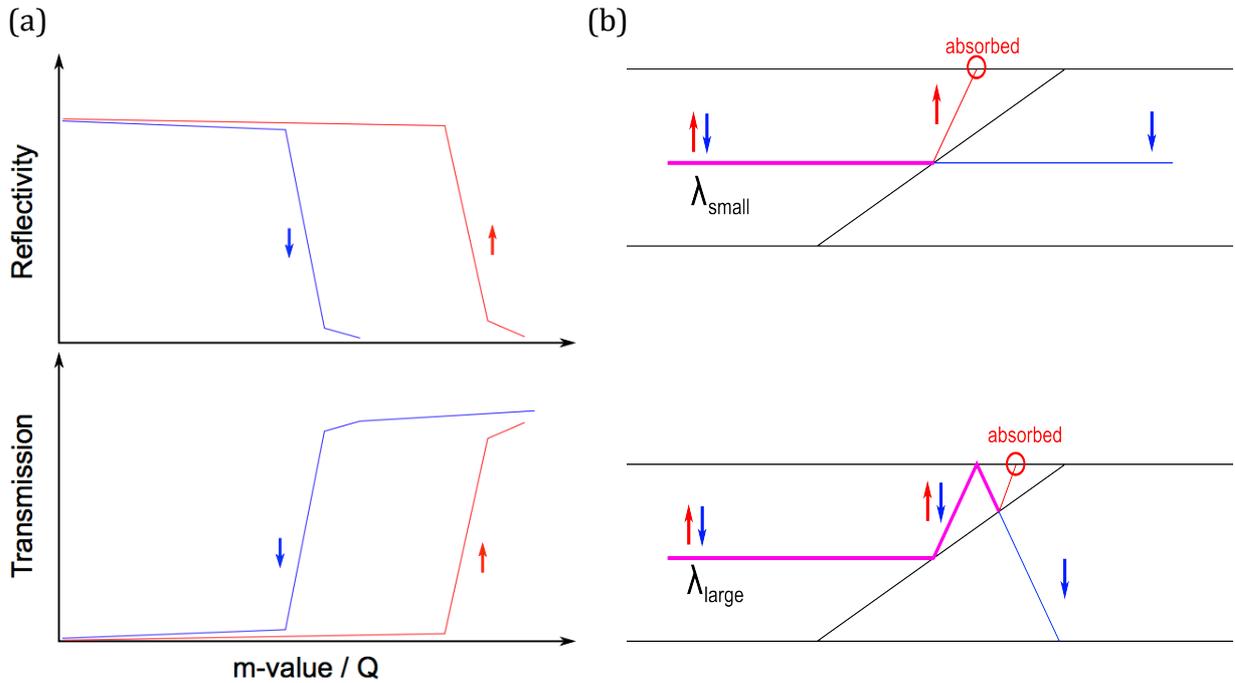

**Fig. 4: (**a) sketch of reflectivity (top) and transmission (bottom) as a function of the m-value of a coating for both spin directions in case of a super-mirror, (b) behavior at a super-mirror for polarization of neutrons for short (top) and long (bottom) wavelength. For short wavelengths a direct transmission of the correct polarization is probable, while the other polarization is reflected away and absorbed at the wall of the neutron guide. In the case of high wavelengths, both polarizations will be reflected and hit the polarizer at a steeper angle, i.e. higher Q, where now again the same situation as in the case of low wavelengths occurs and the undesired wavelength is reflected away.

4.2 Polarizer

The polarizers are based on a construction that is implemented at the KWS-1 SANS instrument in Garching at the MLZ. Here, three cavities with V-shaped super-mirrors serve for a polarization of ca. 95%. The general working principle of such a setup is shown in Fig. 4 and relies on the gap between up and down spin reflectivity and transmission of neutrons at super-mirrors. In such a setup there is a redistribution of low to high divergencies and vice versa [17]. Fig. 5 explains this feature for simple (Z-shaped) and V-shaped cavities. For small wavelength and low divergencies the direct transport of the desired polarization can be achieved. For long wavelengths, a first reflection for high divergencies is likely, and lead to a transformation to higher angles that can be transmitted at the super mirror. The only prerequisite for outgoing neutrons at small divergencies and large wavelengths is the presence of large divergencies beforehand. For the large distance of 2m of the cold source to the first guide segments (the bender) does not allow for the needed large divergencies at the ESS. So, two polarizers for bands of smaller and larger wavelengths are needed for the ESS SANS instrument in order to place the transmission/reflection gaps at the appropriate wavelengths.

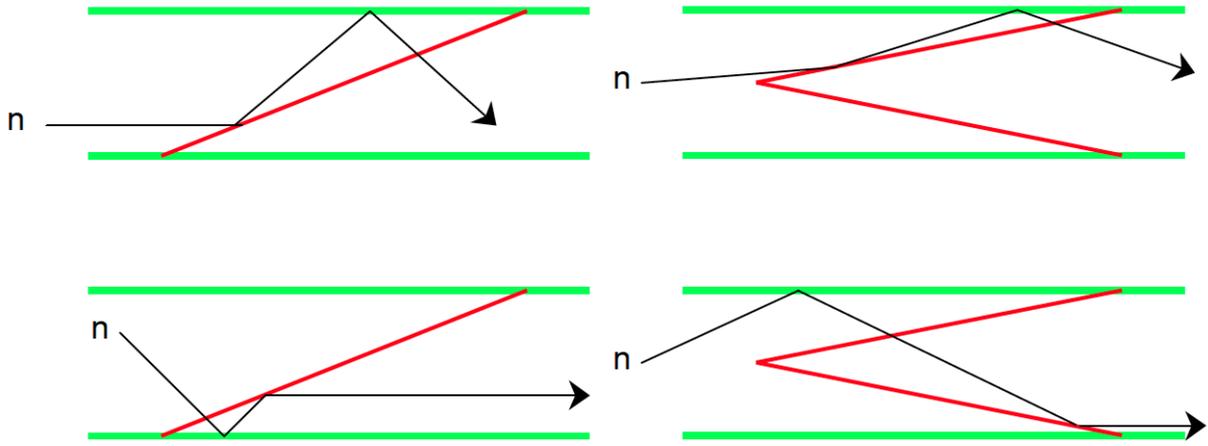

**Fig. 5:** Redistribution of divergencies for a Z-shaped polarizing cavity (left) and a V-shaped polarizing cavity (right).

In order to choose appropriate lengths for both polarizers simulations with different polarizer length were performed (in either case we assumed V-shaped cavities). In Fig. 6a the polarization is plotted against wavelength. Here the complete wavelength band provided by the ESS source of 2 to 20 Å is shown, however in the real instrument a portion of this spectrum will be used and the spectrum will be truncated by choppers. It is visible that the onset of a high degree of polarization shifts to smaller wavelength for an increasing length of the polarizer. The onset of the plateau for all simulated wavelength is shown in Fig. 6b. The onset of the plateau shifts to lower wavelength as already seen in the polarization data. Together with the transmission data shown in Fig. 6c this allows to determine the two different polarizer lengths to provide a high degree of polarization and a high flux simultaneously. For short polarizers the onset of high polarization is at high wavelength and the transmission is constant at 0.5. Therefore a polarizer for a high wavelength band starting at 6 Å should have a length of $L_{Pol}$ = 0.48 m (the angle for the super-mirror is ca. 0.9°, with a coating m = 2.2 for the unwanted polarization). To accommodate for a lower wavelength band starting at values of around 3 Å we have to take into account two different effects: A very high degree of polarization will only be reached at wavelengths slightly higher than the onset and the transmission decreases strongly for very high wavelength above 12 Å. Taking into account both of these factors a length of $L_{Pol}$ = 0.80 m seems prudent (the angle for the super-mirror is ca. 0.54°). This combines an early onset, where the degree of polarization is already above 0.8 at 3.2 Å with a wavelength band with a very high degree of polarization above 0.9 for wavelengths above 6 Å with a high flux, as the transmission only decreases at values above 12 Å. This allows for choosing either high intensities due to the wavelength distribution of the ESS source when choosing to low wavelength as well as a degree of very high polarization for wavelength above 6 Å. This choice is of course dependent on the requirements of each individual experiment.

Two sequential polarizers are currently in discussion, because the imperfection of the polarization (1-P') would be improved according to (1-P') = $(1-P_0)^2$, with $P_0$ being the polarization of a single polarizer. These polarizers have to be well separated (not on a single waver), otherwise the improvement is only (1-P') = ½(1-$P_0$). Computer simulations will answer the question what well separated means in reality. The sharpness of the plateau onset should be improved considerably. Surely, there will be enough space in the instrument for the shorter bender of 0.48m length.

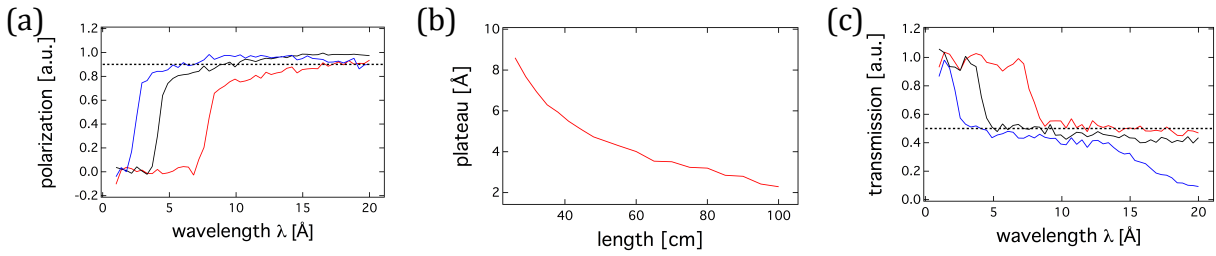

**Fig 6**: (a)Polarization as a function of wavelength for polarization length of 0.26 (red), 0.48 (black) and 0.80 m (blue); dashed line at 0.9, (b) onset of the plateau of high polarization as a function of polarizer length, (c) transmission for different polarizer length as a function of wavelength for polarization length of 0.26 (red), 0.48 (black) and 0.80 m (blue) dashed line at 0.5.

For analyzers we propose $^3$He SEOP (spin exchange optical pumping) cells that are optimized for either small or large angles [18]. Experience exists for the beam lines MARIA and KWS1 and KWS2 MLZ [5]. Limited wavelength bands are not an issue with the proposed SANS instrument, since the default wavelength band is 2 to 7.5Å.

5. Collimation and neutron optics

5.1 Wavelength band selection

The wavelength band will be selected by a set of three choppers placed at 11.5, 14.5, and 21.5 m from the source (Fig. 7). Two co-rotating discs will allow for variable opening times, which is needed for the wide wavelength band mode. The choppers fit in the gaps of the 1m sections of the collimation.

Two different modes for the standard operation are envisaged: The high intensity mode choses the band between 2 and 7.5Å, while the wide Q-range mode skips every second pulse and allows for a band between 2 and 13Å. The first mode is optimized for experiments that explicitly know the interesting Q-range and highest intensities are the priority. The other mode will be used for experiments with less well predictable Q-range (beforehand), and so the maximum dynamic range is the main purpose. Especially, kinetic measurements with growing objects are such an example where the length scales will benefit from that mode of operation.

The flexibility of the chopper settings will allow for any wavelength bands that are reasonable. So, it might be preferential to skip the lowest wavelengths, and focus on a band of 4 to 9.5Å for more classical SANS samples. Apart from this, for focusing on the lowest scattering vectors Q a range of 12 to 17.5Å might be useful. Furthermore, a smaller detector distance of 8m (for the more remote one) would consequently enlarge the usable bandwidth from Δλ = 5.5 to 7.2Å. This relaxation of the time resolution for the time of flight analysis allows for a higher intensity.

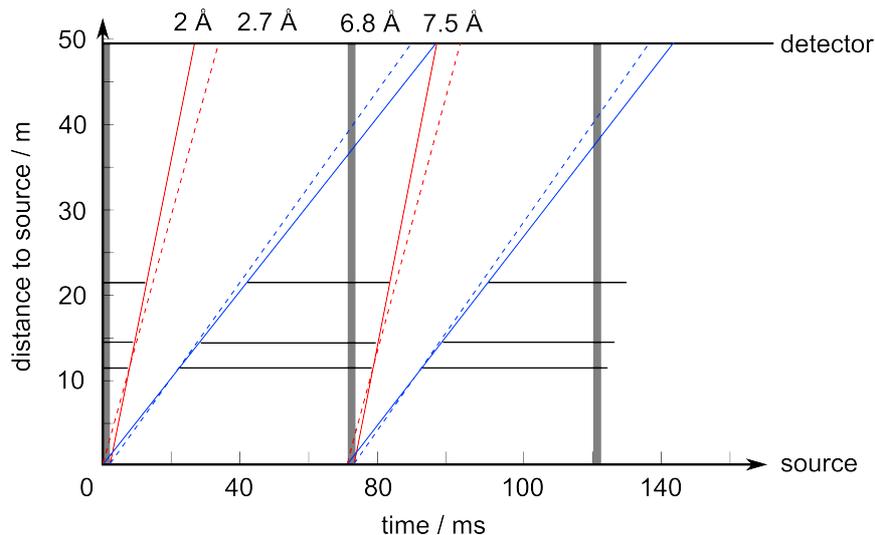

**Fig. 7**: Time-of-flight diagram at a 20m detector setting with a usable wavelength band between 2 and 7.5Å. The choppers are placed at 11.5, 14.5 and 21.5m from the source. The intensity will ramp for the initial 2 to 2.7 and the higher 6.8 to 7.5Å neutrons due to the chopping.

The chopper speed in the normal modes 1 and 2 is 14 and 7 Hz (840 and 420 rpm). The transition times for opening and shutting the beam might be a problem. We propose choppers of 60cm diameter with two co-rotating discs. The typical shutting times are then 0.9 and 1.8ms. If we allow the transition regions of the first and last time frame to overlap, and agree on rejecting these data, then still the central time frames will carry the major intensity. The rejection will neglect small intensities anyway, as ramping at the beginning and the end of the pulse has not yet reached its plateau.

5.2 Collimation

The collimation will be quite narrow for the first 8m for keeping the angular space between different instruments. Here, 1m guide segments will only be moved in or out. On the last 12m, a wider collimation will allow for implementing focusing slits as a VSANS option and a SESANS option for several 10µm large objects.

5.2.1 VSANS

The VSANS method aims at ca. 10 times smaller scattering angles than the classical SANS method. To achieve this, the primary beam must be 10 times smaller (so ca. 3x3mm$^2$) on the detector using focusing methods. There are focusing mirrors and lenses that depict the small entrance aperture on the detector 1:1. Here, the proposed VSANS technique uses multiple converging slits or apertures for the same purpose, beginning with the original cross section of the neutron guide for the complete set of apertures. The sample is always placed at a position where the whole incoming neutron field is still relatively wide (either after the mirror/lenses or in the middle of the first aperture and the detector), and so larger samples are needed for this intensity demanding experiment. Here, for the proposed method, the typical sample size will be 1.5x1.5cm$^2$, and thus not much bigger than for classical SANS. The only disadvantage of the proposed method is the influence of gravity that would result in a wavelength selection in the case of apertures or horizontal slits. Vertical slits would allow for enough flexibility that no wavelength selection takes place. The horizontal slits are nonetheless still in discussion for the combined GISANS/VSANS option when the vertical divergence must be well defined.

The VSANS option will be a focusing multi-slit converging beam collimation that could be operated or removed according to the user needs. The multi-slit concept is a series of N masks made of several slits (typically of 1mm aperture) that reduce the beam size and focus the neutrons on the detector (Fig. 1.12a). The number of masks, their respective positions as well as the number and the size of the slits will be determined by McStas simulations. Both vertical and horizontal slit geometry would be possible for VSANS and respectively GISANS experiments. The total length of the multi-slit collimation system will be between 1 and 3 m and would be located in the front part of the collimator just before the sample position. The expected Q range comprised between several $10^{-4}$ and $10^{-2}$Å$^{-1}$ will permit to extend significantly the classical SANS Q domain, $10^{-3}$ to 1Å$^{-1}$, with a good overlap while keeping for a good spatial resolution.

A feasibility study is currently performed at the LLB with a combination of simulations and real multi-slits TOF experiments on the operating VSANS spectrometer [19]. To achieve a sufficient resolution of the signal on the detector a spatial resolution of 3x3 mm$^2$ is needed.

5.2.2 GISANS
In order to be able to probe surface structures on substrates, a grazing incidence SANS setup can be realized with little additional effort on the proposed SANS instrument, as the essential components, such as a good collimation and high-resolution detector, are already in place. Two horizontal slits that are shaped by the classical rectangular apertures in the collimation will allow for performing experiments with inclining/declining incident angles of up to ±0.2° for a horizontal sample position (the critical angle of D$_2$O vs. air is around 0.5°, and will not be served without declining collimation, that is currently not considered), while vertically the sample will always be turned by the sample stage. The classical Q range will be on the same order of magnitude as for the pure SANS setup without VSANS option (while the combination is still under discussion). McStas simulations will be used to determine an appropriate slit size and position for the desired Q-range and depth resolution, and typical intensities to be expected.

5.2.3 SESANS
The SESANS technique aims at even smaller Q, or larger length scales than VSANS. Typically, several 10µm can be achieved. Technically, the neutron spin is used to encode the different extremely small scattering angles. As for all spin echo methods, a Fourier transformation takes place, that in the case of SESANS delivers the real-space Van Hove correlation function [20]. Since the scattering angles are extremely small, a high resolution detector is needed at the place of the beam stop (preferentially with 10µm resolution), while for the classical SANS angles the scattering intensity can be detected in parallel serving for an extremely wide Q-range simultaneously.

The SESANS option ideally suits to the SKADI instrument for reasons of space in the collimation section. The SESANS option will allow for largest length scales to be measured in the range of several 10µm. The required spatial resolution of the detector for this is around 55x55 µm$^2$.

6. Sample Area

The default sample area will extend by 1m towards the source and a few ten cm towards the detector. From the collimation side, a movable cone will make sure that the vacuum guide extents maximally towards the sample. The detector tube shall provide a relatively large window for largest scattering angles. For huge sample environments, it might be needed to move the detector tube backwards. We propose a motor driven 2m rail system for the whole detector tank. A decoder must make sure to know the real detector position from the tube movement and the inside detector movement to a precision of a few mm. In this way a space of 3m is achieved in the flight direction. For this versatile instrument, a side position at a neutron guide fan would be desirable. In this way a perpendicular space of 3m could be achieved for the sample position, resulting in a maximal space of 9m$^2$ for custom sample environment.

7. Detectors

The general setup of the detectors is planned to be a two-stage system with a first detector at 0.2 of the collimation length and a second one at the full collimation length. The first one will have a 20x20cm$^2$ aperture at low angles to allow neutrons to the second detector for the low Q-range. By virtue of higher distance and better angular resolution the detector at the far end will give a higher Q resolution. Moreover by covering a wide Q range simultaneously this setup accommodates real-time measurements, especially single-shot measurements. The slight overlap between the two detectors will not result in a gap in Q space due to the TOF capability of the instrument.

To achieve these goals and at the same time have a reliable technology in terms of engineering feasibility and endurance we propose a system made up from modular photomultipliers. In combination with a $^6$Li scintillation converter these should allow for count rates in the several ten MHz range over the whole detector, at this point presumably primarily limited by the counting electronics. The pixel resolution of these detectors is ca. 6 mm in both directions, thus there is no need for the Anger approach to improve resolution. New detectors with a pixel resolution of 3x3mm$^2$ are already available, so we propose to use them for the zero-angle detector. Here we will use the highest resolution available at the point of purchase, as substantial improve to values as low as 1.5x1.5mm$^2$ are expected within the next years. Possibly they have to be protected by an Li based attenuator, however this will allow for the measurement of the complete scattering image without impediment by a beam stop. As photomultipliers of these types are already commercially available we do not expect problems during the engineering phase and have requested a prototype for the end of 2014.

For even higher resolution in the low Q area (the primary beam detector) we are looking into single quantum detecting pixel detectors Timepix [21]. These devices can use either Silicon sensor covered with $^6$Li converter [22] or neutron sensitive Micro-Channel-Plate [23]. The pixel size is 55x55 µm$^2$ and sub-pixel resolutions are achievable as well [22-23]. Pixels of these devices can be operated in Time-Of-Arrival mode and hence the devices are directly applicable for Time-Of-Flight applications.

## 8. Instrument performance and Figure of Merit

The determination of the Figure of Merit (FOM) of a SANS instrument is not simple, because an ensemble of samples needs to be considered that represents the whole scenario of samples. Furthermore, information theory [24-25] has only developed a simple expression for a linear detector with a delta-peak signal. This FOM formula reads:

$$\text{FOM} = \frac{I_{max}}{\sigma_Q^2(Q)} \cdot \ln\left(\frac{Q_{max}}{Q_{min}}\right) \qquad (3)$$

The intensity is $I_{max}$, the resolution of the instrument is $\sigma_Q$, and the detector limits are translated to $Q_{min}$ and $Q_{max}$, the limits in physical units. This surely does not represent a SANS experiment, especially not the overall ensemble, where information can be encoded anywhere on the detector.

First of all, we have to assume, that the experiment is adapted to the problem, and the detector really 'looks' where the information is gathered. In this sense, $Q_{min}$ and $Q_{max}$ in the logarithm focus on the signal at Q with a width $\pm\sigma_Q$. Then the logarithm turns to $\sigma_Q/Q$. We arrive then at:

$$FOM \propto \phi_0 \frac{d_E^2}{L_C^2} d_S^2 \cdot \int_{Q_{min}}^{Q_{max}} dQ \left(\frac{Q}{Q_{min}}\right)^{-\alpha} \frac{Q}{\sigma_Q^2(Q)} \frac{\sigma_Q(Q)}{Q} \qquad (4)$$

The circumference of the detector element, where the intensity of equal Q is gathered, is proportional to Q. At this detector element we assume a scattering law $Q^{-\alpha}$, normalized to 1 at the smallest Q. Furthermore, the flux at the sample position is $\phi_0\, d_E^2\, d_S^2 / L_C^2$, with $\phi_0$ being the brilliance transported to the sample, $d_E$ and $d_S$ being the diameters of the entrance and sample apertures, and $L_C$ the collimation distance. The resolution is proportional to [26]:

$$\sigma_Q(Q) \propto \sqrt{Q^2\left(\frac{\Delta\lambda}{\lambda}\right)^2 + 4Q_{min}^2} \qquad (5)$$

with $Q_{min} = (\pi/\lambda)(d_E/L_D)$, and $Q_{max} = (\pi/\lambda)(d_O/L'_D)$, that are connected to the entrance aperture $d_E$ and the outer detector diameter $d_O$ for a symmetric set-up (collimation and detector distance are equal ($L_C=L_D$)) but shorter second detector distance $L'_D$. For the wavelength λ we assume the average pulse wavelength (4.5Å from Fig. 1b at the inner rim), and for Δλ/λ an equally averaged smearing that arises from the pulse length (Δλ/λ = Δt/t = 2.82ms/(Lλ/3.956) = 0.045). For a constant intensity (α=0), we would arrive at:

$$FOM_0 \propto \phi_0 d_S^2 \cdot (\lambda Q_{min})^2 \left(\frac{\Delta\lambda}{\lambda}\right)^{-1} \cdot \ln\left[\frac{\frac{Q_{max}}{Q_{min}} + \sqrt{\left(\frac{Q_{max}}{Q_{min}}\right)^2 + 4\left(\frac{\Delta\lambda}{\lambda}\right)^{-2}}}{1 + \sqrt{1 + 4\left(\frac{\Delta\lambda}{\lambda}\right)^{-2}}}\right] \qquad (6)$$

that can also be expressed in terms of the 'arcsinh(x)' function. In the limit of large solid angle coverage by the neutron detectors and still large $(\Delta\lambda/\lambda)^{-1}$ one would arrive at:

$$FOM_0 \propto \phi_0 d_S^2 \cdot (\lambda Q_{min})^2 \left(\frac{\Delta\lambda}{\lambda}\right)^{-1} \cdot \ln\left[\frac{Q_{max}}{Q_{min}} \frac{\Delta\lambda}{\lambda}\right] \qquad (7)$$

This FOM is important for instruments that try to maximize the detector coverage to quite large solid angles, which is reasonable for aiming at a large dynamic Q-range in single-shot experiments. Alternatively, one might focus on low-Q resolution with focusing elements and relatively small detectors. Then, the FOM has a different limit, namely:

$$FOM_0 \propto \phi_0 d_S^2 \cdot (\lambda Q_{min})^2 \cdot \frac{1}{2}\left[\frac{Q_{max}}{Q_{min}} - 1\right] \qquad (8)$$

In this limit, the wavelength spread is not important, and only the relatively small detector dimensions strongly contribute. One usually tries to compensate the smallest minimum Q values by large samples for large intensities.

For faster decays of the ensemble of scattering laws one would obtain in explicit cases of α=1 the following:

$$FOM_1 \propto \phi_0 d_S^2 \cdot (\lambda Q_{min})^2 \frac{1}{2}\left[\operatorname{arctanh}\left(\left(\frac{1}{4}\frac{Q_{max}^2}{Q_{min}^2}\frac{\Delta\lambda^2}{\lambda^2}+1\right)^{-1/2}\right) - \operatorname{arctanh}\left(\left(\frac{1}{4}\frac{\Delta\lambda^2}{\lambda^2}+1\right)^{-1/2}\right)\right]$$

$$\rightarrow \phi_0 d_S^2 \cdot (\lambda Q_{min})^2 \frac{1}{2} \cdot \begin{cases} -\ln\left(\frac{1}{4}\frac{\Delta\lambda}{\lambda}\right) - 2\frac{Q_{min}}{Q_{max}}\frac{\lambda}{\Delta\lambda} & \text{for large detectors} \\ \frac{Q_{max}}{Q_{min}} - 1 & \text{for high resolution} \end{cases} \qquad (9)$$

and for α=2:

$$FOM_1 \propto \phi_0 d_S^2 \cdot (\lambda Q_{min})^2 \frac{1}{4}\left[\sqrt{\frac{\Delta\lambda^2}{\lambda^2}+4\frac{Q_{min}^2}{Q_{max}^2}} - \sqrt{\frac{\Delta\lambda^2}{\lambda^2}+4}\right]$$

$$\rightarrow \phi_0 d_S^2 \cdot (\lambda Q_{min})^2 \frac{1}{2} \cdot \begin{cases} 1 - \frac{1}{2}\frac{\Delta\lambda}{\lambda} - \frac{Q_{min}^2}{Q_{max}^2}\frac{\lambda}{\Delta\lambda} & \text{for large detectors} \\ \frac{Q_{max}}{Q_{min}} - 1 & \text{for high resolution} \end{cases} \qquad (10)$$

The comparison for the approaches with different α is summarized in Fig. 8. For α=0, the detector is homogenously considered, and the Figure-of-Merit grows logarithmically with large detector coverage, similarly to the original Formula for an isolated delta peak. Contrarily, for α=1 and 2, the $FOM_{1/2}$ saturates at large detector coverage, because the small Q are emphasized too strongly. From a practical view, the $FOM_0$ seems to be the suitable function for evaluating a SANS instrument.

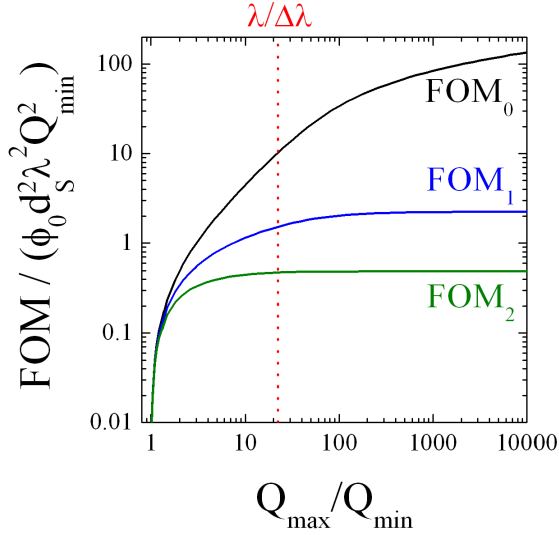

**Fig. 8**: Analytical Figures-of-Merit (FOM$_\alpha$) for differently decaying weights $Q^{-\alpha}$ as averaged scattering functions. For $\alpha=0$, the FOM$_0$ is logarithmically growing with large detector coverage. For $\alpha=1$ and 2, the FOM$_{1/2}$ is saturated at large detector coverage due to the high emphasis at low Q.

8.1 Computer simulations

For computer simulations we used the McStas package from the Risø National Laboratory [15-16], and developed our own sample and detector components. The sample component allowed for analytical and tabulated scattering laws, and distinct delta peaks at constant Q in the range of $10^{-4}$, $10^{-4+1/3}$, $10^{-4+2/3}$, … $10^{1/3}$Å$^{-1}$ with the experimental scattering intensity being constant. Incoherent background and absorption can be simulated as well. The sample component serves for correct absolute intensities. Multiple scattering effects can be switched on, but have not been used for the benchmarking here. The sample component puts emphasis, that mainly small angle scattering and a balanced transmission are simulated. So the intensity is distributed equally on logarithmic Q-scale. Due to incoherent scattering, only a few neutron paths are directed backwards and are usually neglected by the following component.

The detector component records the scattered neutrons on two square area detectors. The first detector has a hole in the center. The second one carries a small beam-stop that records the beam center and intensity (for transmission measurements). All neutrons are binned according to a given pixel size in 2d-space, and to time slices according to the desired TOF-analysis. At the end of the simulation, the radial averaging is carried out, and intensities are binned according to the same Q. Pure count rates and absolute calibrated intensities are written to different files.

These two components proved to be highly valuable for fast simulations of SANS experiments. Around the sample/detector we programmed a loop of 1000 repetitions to account for the different possibilities of scattering. With 8m as the collimation and maximum detector distance, a single simulation with acceptable statistics was performed in 10 mins on a MacBook Pro. For 20m, the simulation time was ca. 1 hour.

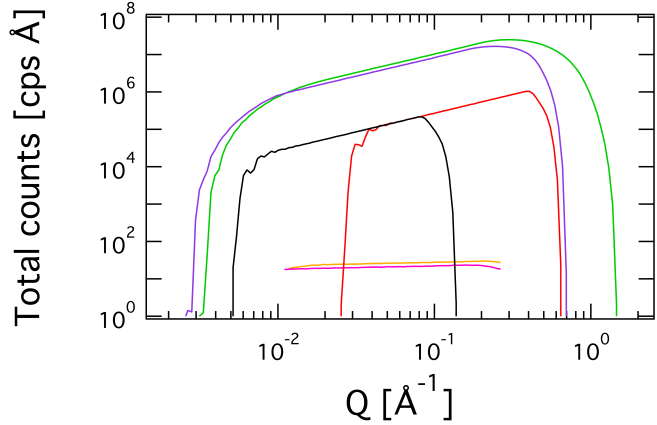

**Fig. 9**: Intensities of 1mm water sample (1x1cm$^2$) for instruments SKADI (green, lilac) and D22 (black, red). The divergence was identical (proposed SANS: 3x3cm$^2$ diaphragm at 4.8m collimation and D22: 4x5.5cm$^2$ at 8m). The detectors were placed at 8 (black) and 1.6m (red). For SKADI the wavelength bands of 2 to 9.2 (green) and 4 to 11.2Å (lilac) were chosen. The pure intensity ratios are between 20 and 29 at each given Q value (lower lines, orange, pink).

8.1.1 Bare intensity gain

For the bare intensity simulations, we compared the proposed SANS instrument at the ESS with the D22 instrument at the ILL [27]. The complete McStas code for D22 was obtained from E. Farhi. The proposed SANS instrument relied on current McStas components published on the web. The results are depicted in Fig. 9. First of all, the proposed SANS instrument will simultaneously cover a wide Q-range from ca. $3 \times 10^{-3}$ to 1Å$^{-1}$ for the 8m and 1.6m detector distances. The used wavelength band ranged from 2 to 9.2Å (alternatively from 4 to 11.2Å). For D22, two distinct detector settings were needed. The obtained intensity ratios were in the range of 20 to 29 – neglecting the different detector coverage.

8.1.2 Simulated Figure of Merit

For the correct Figure of Merit comparison of the two instruments, the sample produced Debye-Scherrer rings of infinite sharpness that is only changed by the instrumental resolution. The peaks were placed at Q = $10^{-4}$, $10^{-4+1/3}$, $10^{-4+2/3}$, ... $10^{1/3}$Å$^{-1}$ as before. From this experiment, the maximum count rates of the different peaks are determined and the corresponding peak width. The corresponding FOM of this simulation experiment reads then:

$$\mathrm{FOM} = \sum \frac{I_{\mathrm{peak}}}{\sigma_Q(Q)} \quad (11)$$

The comparison between D22 and SKADI resulted in a gain of the FOM by a factor of 35.8 and is quite independent of the exact settings of the instrument (see also Fig. 10). This especially supports this concept of the FOM.

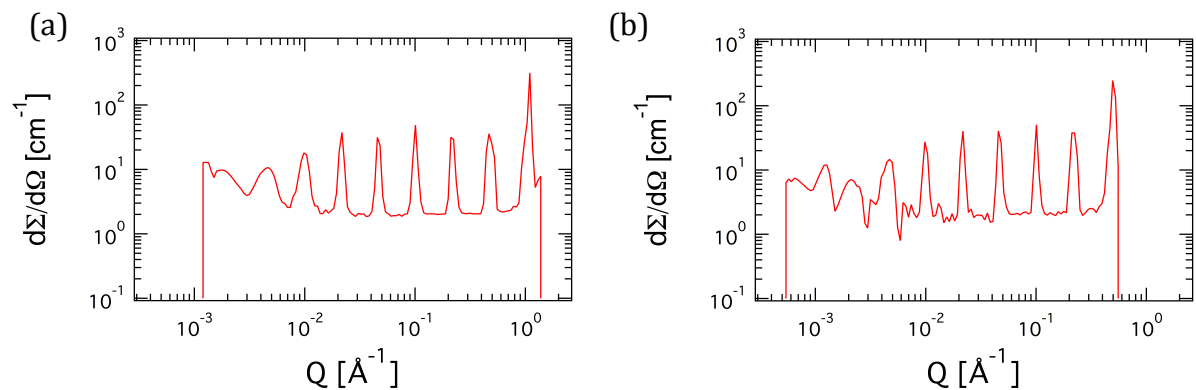

**Fig. 10**: (a) Resolution simulation for the proposed SANS instrument, calculated back to absolute intensities. The detectors are placed at 8m and 1.6m. The used wavelength band was 2 to 15Å without choppers (simulated skipping every 2nd pulse). Only a single pulse was simulated, so there is no overlap of slow and fast neutrons from consecutive pulses. (b) Here, the detectors are placed at 20m and 4m. The used wavelength band was 2 to 13Å.

9. Summary

A new SANS instrument for the ESS is proposed. This instrument will make ideal use of the 2.8ms pulses at 14Hz for small Q of ca. $10^{-3}$Å$^{-1}$ by a long collimation and detector distance of 20+20m. So the instrument with bender and polarizer is (9.5+20+20)m ≈ 50m long. The bender ideally suppresses fast neutrons. The polarizer provides polarized neutrons for studying magnetic structures and subtracting incoherent background correctly. An analyzer will allow for distinguishing spin-flip and non-spin-flip scattering. SKADI will host VSANS and SESANS options that increase the resolution of the instrument towards μm and tens of μm, respectively. In the 20m long detector tube, two 1m$^2$ area detectors will span a dynamic Q-range of ca. 1000. The performance of the new SANS instrument against the instrument D22 at ILL has an expected intensity gain of 20 to 29, while the figure of merit gain is even ca. 36 due to the better resolution. This will open new perspectives for fast kinetics, weakly scattering samples, and multi-dimensional contrast variation studies.

Acknowledgements

We want to thank for fruitful discussions: Andrew Jackson (ESS), Richard Hall-Wilton (ESS), Kalliopi Kanaki (ESS), Richard Heenan (ISIS), Kenneth Andersen (ESS).